\documentclass[sn-mathphys,Numbered]{sn-jnl}

\usepackage{graphicx}%
\usepackage{multirow}%
\usepackage{amsmath,amssymb,amsfonts}%
\usepackage{amsthm}%
\usepackage{mathrsfs}%
\usepackage[title]{appendix}%
\usepackage{textcomp}%
\usepackage{manyfoot}%
\usepackage{booktabs}%
\usepackage{algorithm}%
\usepackage{algorithmicx}%
\usepackage{algpseudocode}%
\usepackage{listings}%
 \usepackage{multirow}
 \usepackage{graphicx}
 \usepackage[table,xcdraw]{xcolor}

\begin{document}

\title[Article Title]{ Roadmapping the Next Generation of Silicon Photonics}


\author*[1]{\fnm{Sudip} \sur{Shekhar}}\email{sudip@ece.ubc.ca}

\author[2]{\fnm{Wim} \sur{Bogaerts}}

\author[1]{\fnm{Lukas} \sur{Chrostowski}}

\author[3]{\fnm{John E.} \sur{Bowers}}

\author[4]{\fnm{Michael} \sur{Hochberg}}

\author[5]{\fnm{Richard} \sur{Soref}}

\author*[6]{\fnm{Bhavin J.} \sur{Shastri}}\email{bhavin.shastri@queensu.ca}

\affil*[1]{\orgdiv{Department of Electrical \& Computer Engineering}, \orgname{University of British Columbia}, \orgaddress{\street{2332 Main Mall}, \city{Vancouver}, \postcode{V6T1Z4}, \state{BC}, \country{Canada}}}

\affil[2]{\orgdiv{Department of Information Technology}, \orgname{Ghent University - IMEC}, \orgaddress{\street{Technologiepark-Zwijnaarde 126}, \city{Ghent}, \postcode{9052}, \country{Belgium}}}

\affil[3]{\orgdiv{Department of Electrical \& Computer Engineering}, \orgname{University of California Santa Barbara}, \orgaddress{ \city{Santa Barbara}, \postcode{93106}, \state{CA}, \country{USA}}}

\affil[4]{\orgname{Luminous Computing}, \orgaddress{\street{4750 Patrick Henry Drive}, \city{Santa Clara}, \postcode{95054}, \state{CA}, \country{USA}}}

\affil[5]{\orgdiv{College of Science and Mathematics}, \orgname{University of Massachusetts Boston}, \orgaddress{\street{100 William T. Morrissey Blvd.}, \city{Boston}, \postcode{02125}, \state{MA}, \country{USA}}}

\affil[6]{\orgdiv{Department of Physics, Engineering Physics \& Astronomy}, \orgname{Queen's University}, \orgaddress{\street{64 Bader Lane}, \city{Kingston}, \postcode{K7L3N6}, \state{ON}, \country{Canada}}}

\abstract{Silicon photonics has developed into a mainstream technology driven by advances in optical communications. The current generation has led to a proliferation of integrated photonic devices from thousands to millions—mainly in the form of communication transceivers for data centers. Products in many exciting applications, such as sensing and computing, are around the corner. What will it take to increase the proliferation of silicon photonics from millions to billions of units shipped? What will the next generation of silicon photonics look like? What are the common threads in the integration and fabrication bottlenecks that silicon photonic applications face, and which emerging technologies can solve them? This perspective article is an attempt to answer such questions. We chart the generational trends in silicon photonics technology, drawing parallels from the generational definitions of CMOS technology. We identify the crucial challenges that must be solved to make giant strides in CMOS-foundry-compatible devices, circuits, integration, and packaging. We identify challenges critical to the next generation of systems and applications - in communication, signal processing, and sensing. By identifying and summarizing such challenges and opportunities, we aim to stimulate further research on devices, circuits, and systems for the silicon photonics ecosystem.}


\maketitle

\section{The Generational Roadmap}\label{sec1}

\begin{figure}[h]%
\centering
\includegraphics[width=0.9\textwidth,trim=0.2cm 0.2cm 0.2cm 0.2cm, clip]{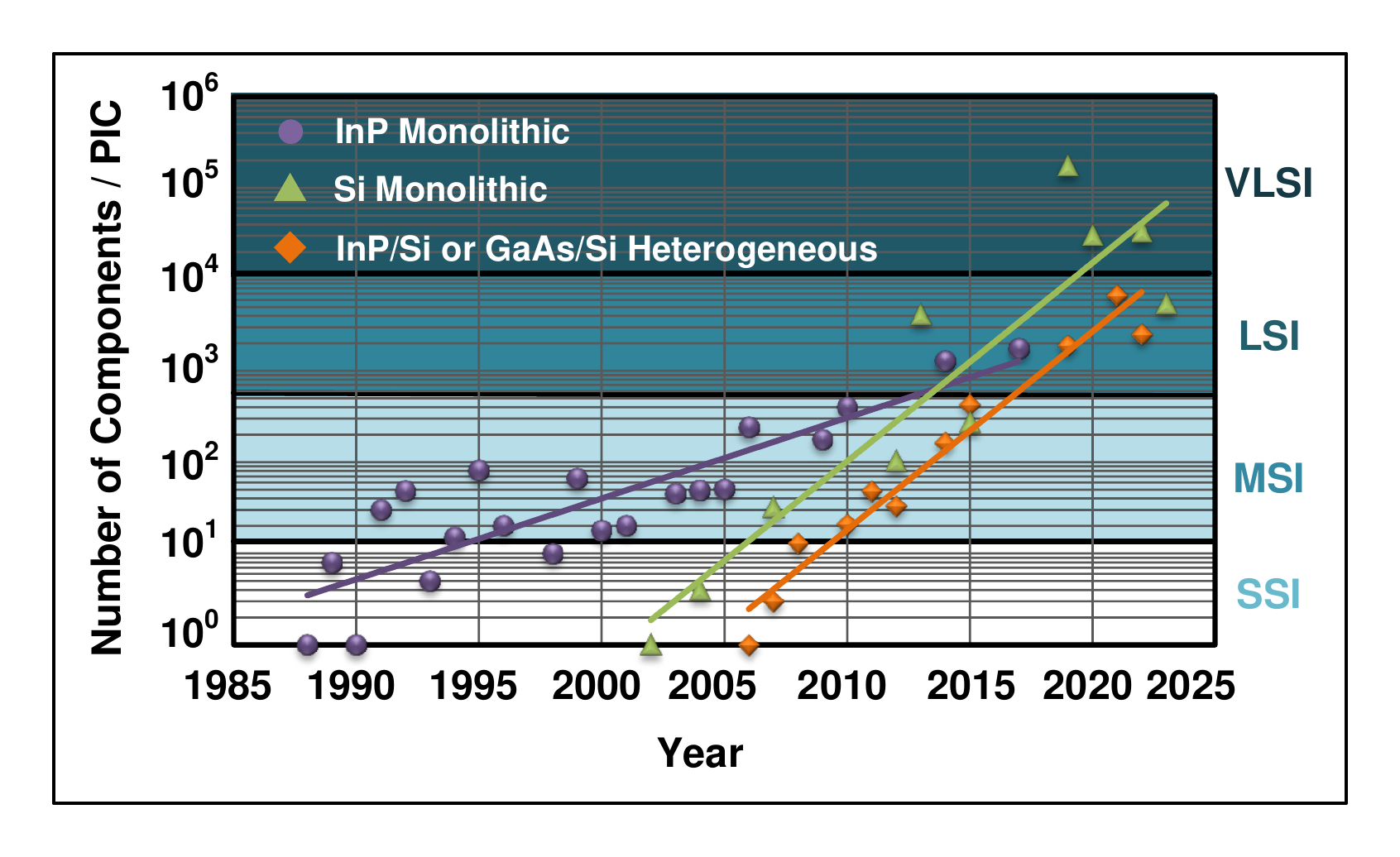}
\caption{Timeline for the number of components on a silicon photonic integrated circuit (PIC) over generations of small-scale, medium-scale, large-scale and very-large-scale integration (SSI, MSI, LSI, VLSI, respectively). A component is a unit cell that is combined with other unit cells to build a circuit, such as a waveguide, directional coupler, heater, grating coupler, etc.  Heterogeneous silicon photonics lags hybrid by approximately two years. For comparison, data for InP-based integrated photonics is also shown. In general, the higher the number of high-speed modulators, the more challenging the scaling. The figure is adapted from~\cite{Margalit21, Khanna17}.
}\label{fig1}
\end{figure}

\begin{figure}[h]%
\centering
\includegraphics[width=0.9\textwidth,trim=0.2cm 0.5cm 0.2cm 1.7cm, clip]{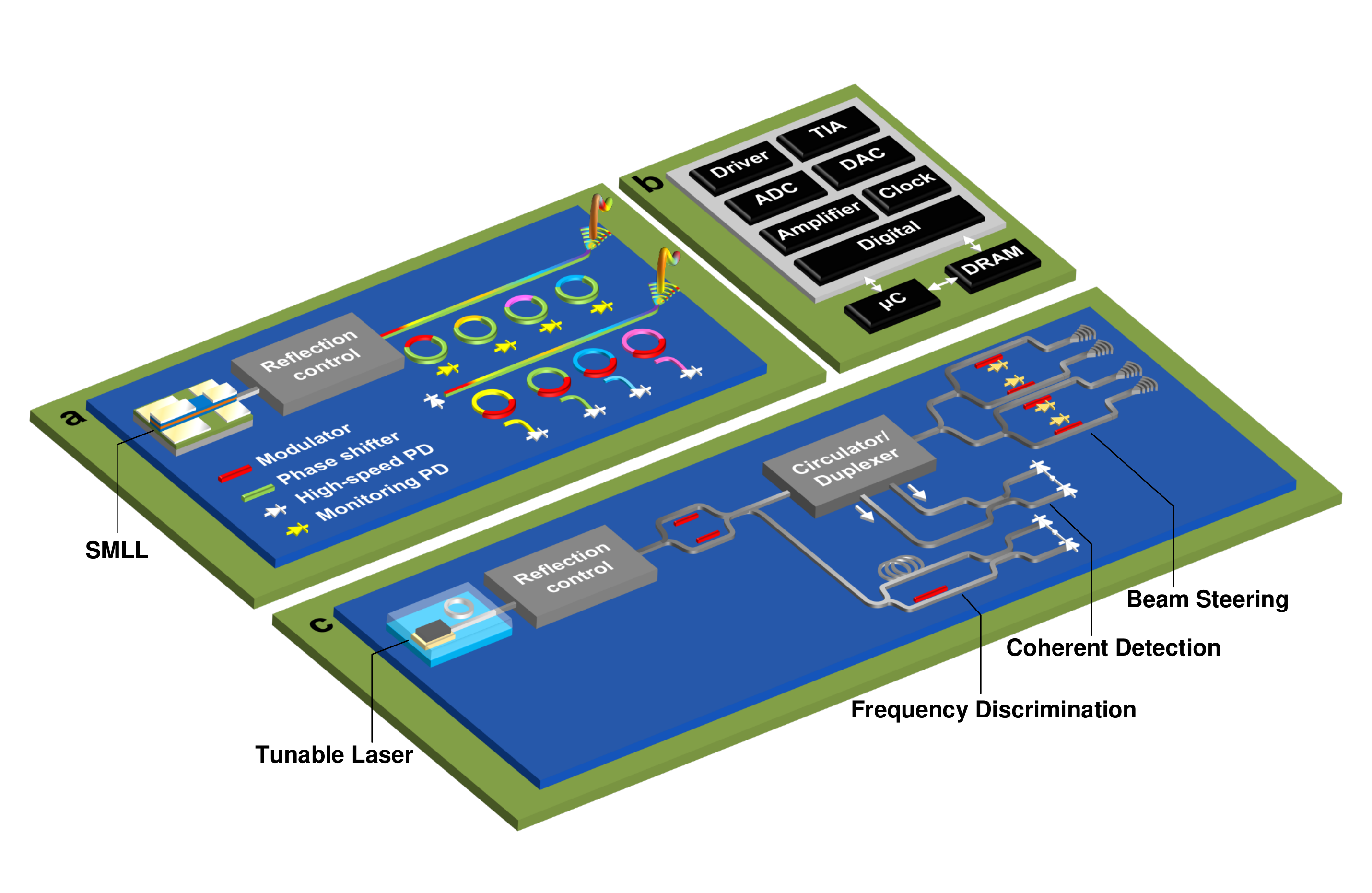}
\caption{Illustrative renditions of LSI silicon photonic systems capturing current and future technologies: a. WDM Transceiver: A semiconductor mode-locked laser (SMLL) provides multi-wavelength continuous-wave (CW) light to an array of compact, WDM-capable modulators and filters. Reflection control circuits limit back reflections into the laser. High-speed photodetectors (PDs) carry out the O/E conversion. b. The electrical current is then amplified by transimpedance amplifiers (TIAs) and limiting amplifiers. Analog-to-digital converters (ADCs) are used to digitize the signal for further digital signal processing (DSP). Monitoring PDs are used for control and stabilization of wavelength, phase shift and temperature. Digital-to-analog converters (DACs) and drivers are used for E/O modulation of the digital signal. Dynamic random-access memory (DRAM) provides large memory access. Micro-controllers ($\mu$C) may be used to offload some of the digital processing as well. c. LIDAR: A tunable laser provides frequency chirped light to a network of phase shifters, circulators/duplexers and coherent frontend for homodyne/heterodyne frequency-modulated CW (FMCW) ranging and detection. Beamsteering is done using optical phase arrays (OPAs) or focal-plane arrays (FPAs). Delay line interferometers aid in calibrating the received beat frequencies and support chirp linearization by directly controlling the tunable laser or a modulator and various forms of error correction through DSP.
}\label{figLSI}
\end{figure}

Fig. 1 maps the evolution of silicon photonics~\cite{Khanna17,Margalit21}. Silicon-based photonic integrated circuits (PICs) were introduced in 1985~\cite{Soref85} and low-loss waveguides in a thick silicon on insulator (SOI) process demonstrated in 1991-92~\cite{Schmidtchen, Weiss}. Various optical devices were next demonstrated~\cite{Bestwick}, and soon, silicon photonics was in the small-scale integration (SSI) era - with 1-to-10 components on a PIC. They included demonstrations of high-speed pn junction modulators~\cite{Liu04, Liao05, Liao07} and photodetectors (PDs)~\cite{Dehlinger04, Ahn07, Vivien07, Yin07}, as well as heterogeneous integration of a III-V laser to a silicon PIC~\cite{Fang06}. The next era ushered in the commercial success of silicon photonics. With 10-to-500 components on a PIC, this medium-scale integration (MSI) era saw successful demonstration and adoption of Mach-Zehnder modulator (MZM) in intensity-modulated direct-detect (IMDD) transceivers within data centers - both single- wavelength~\cite{Luxtera06} and multi-wavelength~\cite{Luxtera07, Luxtera16, Alduino4x10, Jones19}. Microring-modulator (MRM)-based IMDD transceivers (see Fig.~\ref{figLSI}a) demonstrated the multiplexing and energy-efficiency benefits of PIC technology~\cite{Ayar4L, Ayar8L, Fathololoumi22}. Coherent transceivers in silicon photonics/electronics platforms proved that the technology could compete in performance with their LiNb$\rm{O_3}$ photonic and III-V electronic counterparts~\cite{Milivojevic13, DoerrOFC14, Ahmed19}. Besides communications, silicon photonics also found new applications such as evanescent-field biosensors~\cite{iqbal2010}. Silicon photonics is now embarking on the next era of large-scale integration (LSI) - towards 500-to-10000 components on the same chip. Applications for LSI include LIDAR (see Fig.~\ref{figLSI}b)~\cite{Poulton20, Riemensberger20, Zhang21, Zhang22, Rogers21, Poulton22}, image projection~\cite{Raval18}, photonic switching~\cite{Seok19}, photonic computing~\cite{Ramey20,Huang21,Bhavin21, Bandyopadhyay22,Ashtiani22}, programmable circuits~\cite{Wim20}, and multiplexed biosensors~\cite{Reed22}. Even VLSI ($>$ 10000 components) prototypes have now been demonstrated~\cite{Seok19, Zhang22, Poulton22}. In the field of communication, which has been the essential market driver for silicon photonics, silicon photonics has transformed from a challenger technology in the SSI era to arguably a dominant technology in the MSI era for intra-, and inter-datacenter interconnects, and it is poised to become the incumbent technology in the LSI era. For co-packaged optics (CPO) to succeed, high-performance computing to scale~\cite{Fathololoumi22}, and disaggregated computing to become a reality~\cite{Michelogiannakis23}, silicon photonics will be pivotal. 

\section{Silicon Photonics: Technology Perspective}\label{sec2}

Through the generations of CMOS process development, many materials were added to silicon to reduce the Power, improve the Performance, and shrink the Area - often called the PPA metrics. The additions include Al and Cu for metal traces, Ge for inducing strain and enabling heterojunction BJTs, and silicon nitride (SiN) for passivation and diffusion barriers. The CMOS R\&D budgets and commercial markets are orders of magnitude larger than for silicon photonics, so it is natural for silicon photonics foundries to learn from and adopt the innovations from CMOS processes. Hence, we have seen a similar trend in silicon photonics process development. Besides p/n dopants for high-speed modulation, two materials that are now natively supported by several foundries are (1) Ge high-speed photodetectors~\cite{Luxtera18}, and (2) SiN to expand the wavelength range, enable higher optical power, and support waveguides with lower loss and better phase control in interferometric devices~\cite{Bauters13}. 

Shrinking the area will be a key focus for the next decade of silicon photonics process development for the LSI and VLSI era. In reality, the biggest density limitations rarely come from device size; the spacing between waveguides to eliminate crosstalk is much larger than the size of the actual waveguides. For radio-frequency (RF) devices, spacings between active elements - which are microns in critical dimension - are often in the hundreds of microns, to eliminate RF crosstalk. Shrinking these `blank spaces’ requires very detailed systems-level simulation and aggressive multi-physics modeling, and will be at the heart of making chips smaller, cheaper, and higher density. The passives themselves are generally limited in size reduction by the index contrast and the operating wavelength of 1-2$\mu m$. There is still some headroom with the use of inverse design techniques to shrink passive building blocks, but the waveguide itself cannot really shrink much below today’s 400-500~nm width for silicon platforms. However, significant scaling is still possible in the optical I/O couplers and high-speed modulators. For coupling to optical fibers, V-grooves with edge couplers provide low-loss, easy-to-package connectivity at the cost of a considerable chip area. Edge couplers without V-grooves are smaller but require more precise active alignment and surface treatment (polishing, dicing), thereby increasing cost. Multicore fibers are an attractive solution for efficiently using limited photonic beachfront around the edges of a chip~\cite{Lindenmann15}. The main alternative coupling approach is through grating couplers, which are compact, provide the flexibility of positioning on the chip surface, enable wafer level testing, and can also be realized with low insertion loss (IL), but suffer from polarization and temperature sensitivity and lower optical bandwidth~\cite{Cheng20}. Passive alignment packaging techniques, such as photonic wire bonding (PWB)~\cite{Blaicher20}, offer an attractive potential alternative. Using computer vision and automation, PWBs can be fabricated in polymer photoresist through two-photon absorption between two coupling sites allowing up to 30~$\mu$m of offset. Simple alignment markers are used to locate the coupling sites, and the sites do not require strict pitch or large footprints, thereby providing a passive-aligned, low-loss, scalable port count. In another passive alignment technique for pluggable connection, the complexity and accuracy requirement can be moved from fiber assembly to wafer-level manufacturing, where a fiber-receptor die can be flip-chip integrated to the silicon photonic die with a glass spacer~\cite{Terramount20}. Using a combination of V-grooves and mirrors in the fiber-receptor die, and mirrors and surface couplers on the silicon photonic die, a confocal imaging assembly tolerant to $>10~\mu$m relative displacements of the two dies can be realized, providing a passive-aligned, low-loss, scalable port-count and pluggable connector~\cite{Terramount20}. More reliability studies for these passive alignment-based assemblies will be helpful for broad adoption.

\subsection{E/O Modulation}\label{EOMod}
The central quest for the next decade in shrinking photonic chips and thus increasing density is to find the elusive ‘ideal’ modulator in silicon photonics - small in length (L), requiring a small drive voltage to incur a $\pi$ phase shift ($V_{\pi}$), offering low propagation loss ($\alpha$) and IL, and for several applications, highly linear and with large -3dB E/O bandwidth (BW)~\cite{Taghavi22}. Also, this modulator is preferably a phase shifter, as this enables higher-order coherent modulation formats. 

\subsubsection{High-Speed Modulators}\label{Mod}


\begin{figure}[h]%
\centering
\includegraphics[width=0.7\textwidth,trim=0.02cm 0.02cm 0.02cm 0.02cm, clip]{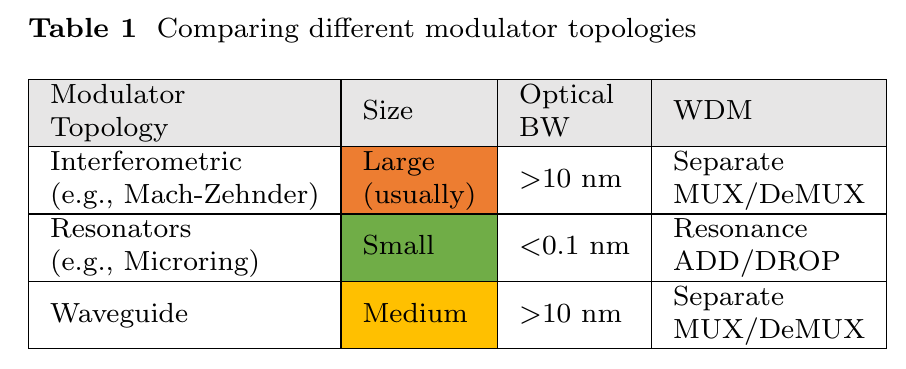}
\label{tab:modtop}
\end{figure}

\begin{figure}[h]%
\centering
\includegraphics[width=0.99\textwidth,trim=0.02cm 0.02cm 0.02cm 0.02cm, clip]{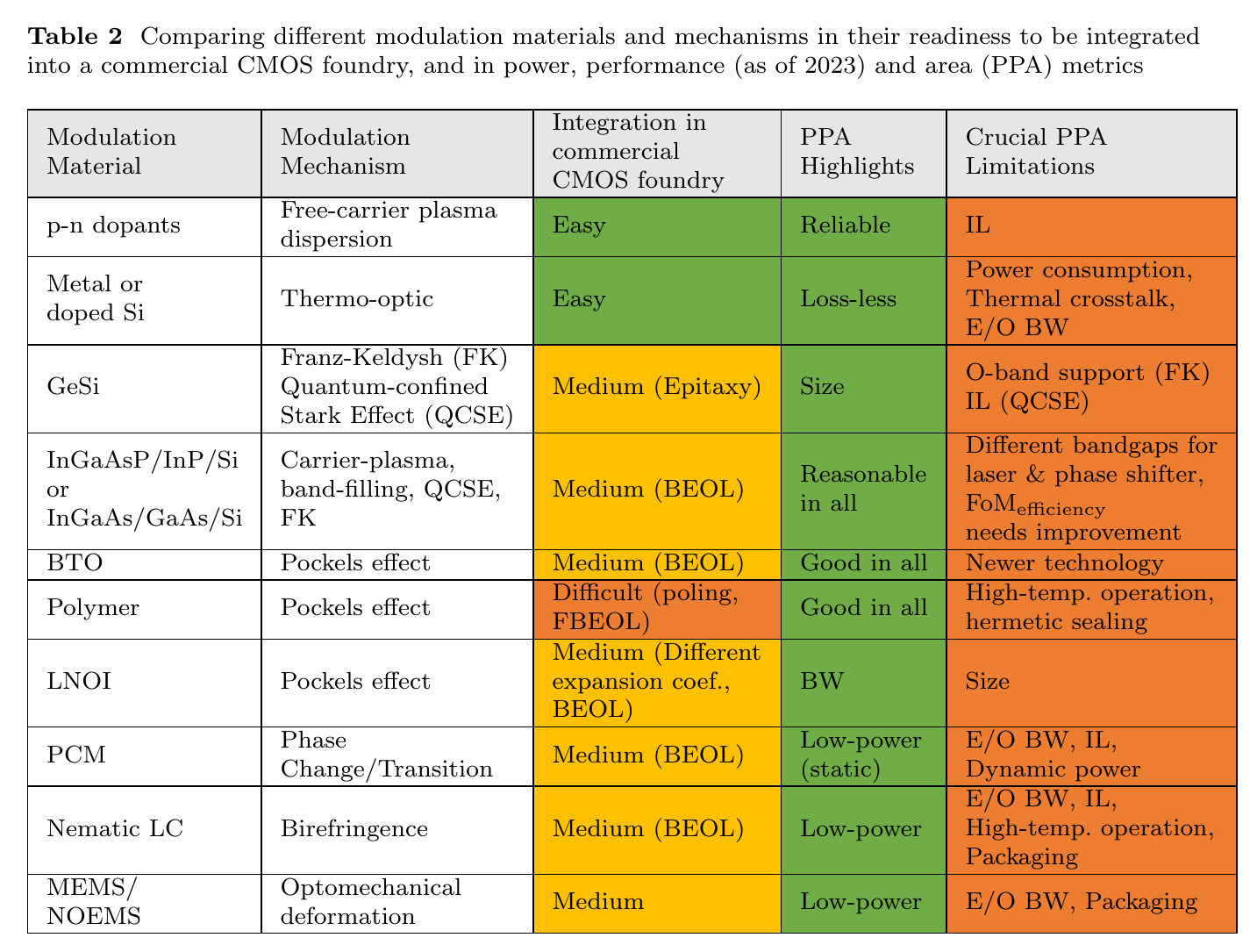}
\label{tab:modPPA}
\end{figure}

The often-used efficiency figure-of-merit (FoM$\rm{_{efficiency}}$) of waveguide-based modulators (Table~\ref{tab:modtop}) is $\alpha V_\pi L$. For MRMs, which are very compact, the loss due to $\alpha$ becomes less critical~\cite{Sun19}, and a better FoM$\rm{_{efficiency}}$ inculcates the IL and $V\rm{_{pp}}$ (the peak-to-peak voltage swing for a specific optical modulation amplitude  or OMA). All modulators suffer from a tradeoff between FoM$\rm{_{efficiency}}$ and E/O BW~\cite{Yu14, Sun19, Taghavi22}. Finally, the power consumed in the driver depends on the modulator impedance as seen by the driver. A resistive impedance (a terminated traveling-wave modulator) consumes static (DC) and dynamic (AC) power, whereas a high impedance (capacitive) consumes primarily dynamic power. A high IL also is a proxy for higher power consumption since the laser power needs to be increased to compensate for the losses.   

Besides the topology, the PPA metrics for a modulator depend on the material and mechanism used for modulation. Table~\ref{tab:modPPA} shows the different materials used for modulators in various silicon photonic processes. p-n dopants utilizing the free-carrier plasma dispersion are natively available in all commercial silicon photonic foundries today, supporting 60~GHz or even higher E/O BW. Currently, the commercial market is dominated by such devices, in the form of traveling-wave MZM modulators~\cite{Murray22}. Plasma dispersion in Si leads to mediocre FoM$\rm{_{efficiency}}$, with high IL for average OMA. Carrier accumulation allows for shorter MZMs, but with BW limitations~\cite{Wu13, Talkhooncheh22}. When implemented as MRMs, the devices are much smaller, but IL and OMA remain suboptimal to support LSI/VLSI ICs. 

With Ge PDs already supported by most commercial silicon photonics foundries, various teams have attempted to use GeSi, a related but not identical technology, to implement a better modulator. GeSi electro-absorption modulators (EAM) based on the Franz-Kelydysh effect can operate in the C/L band at high E/O BW. However, they are generally not optically broadband since they use band-edge modulation for absorption. For O-band operations, modulators utilizing the quantum-confined Stark effect (QCSE) still suffer from a large IL~\cite{Srinivasan21}. While there have been multiple academic and commercial efforts in this space, it is unclear whether these modulators will find their way into future generations of commercial devices.

Heterogeneous integration of modulator technologies -- InP, thin-film LiNbO$_3$ on insulator (LNOI), or thin-film BaTiO$_3$ (BTO) -- with Si can be done using die-to-die, die-to-wafer, or wafer-to-wafer direct (molecular) bonding or adhesive-assisted bonding. Die-to-wafer bonding provides the flexibility of using known-good dies, increasing yield. Wafer-to-wafer bonding remains expensive because the size mismatches between the SOI acceptor wafers (200~mm or 300~mm) and the modulator donor wafers (150~mm or smaller) lead to wastage.

Integration proximity of the (dissimilar) materials in direct bonding facilitates superior optical coupling and heat transportation between them~\cite{Liang10}. However, very smooth and clean surfaces are required. Chemical mechanical polishing (CMP) procedures, already used in high-volume manufacturing (HVM) for heterogeneous direct bonding of InP to Si for lasers, must be optimized for a scalable modulator integration pathway. Annealing is needed for strong molecular bonding and outgassing, but the pre-processed SOI wafer significantly restricts the annealing temperature. Therefore, “low-temperature” annealing at $<$350$^{\circ}$C is usually used, but this necessitates developing custom outgassing techniques and direct bonding recipes that require extensive resources to improve yield. Mismatches in the coefficient of thermal expansion (CTE) must also be minimized. Surface topography requirements can be relaxed, and bonding strengthened using an intermediate adhesive~\cite{Weigel18, WangLN22,Roelkens06} but thermal dissipation, long-term stability, optical power handling, and drift properties need to be studied further~\cite{Mookherjea23}.

Heterogeneous integration of InP to CMOS has already been of interest for electronics~\cite{Royter09} and photonics. For photonics, it has paved the path for laser integration in HVM of IMDD transceivers~\cite{Jones19} and is being used for SOA integration~\cite{Fathololoumi22}. In light of these integration efforts, InP/Si modulators remain promising~\cite{Tang12, Han17, Hiraki21}. InP/Si EAMs for C/L/O bands have been demonstrated and are available in at least one silicon photonics foundry.
Both O$_2$ plasma-assisted~\cite{Tang12} or SiO$_2$ covalent direct wafer bonding techniques~\cite{Liang10} have been adopted. However, the bandgaps required for optimal laser and modulator operation differ, significantly complicating the heterogeneous integration. The state-of-the-art in both FoM$\rm{_{efficiency}}$ and E/O BW needs to be remarkably improved for InP modulators to be popularly adopted as the integration of choice for the next generation of silicon photonics. In general, compatibility to Ge PD processing~\cite{Eltes19} and reflow~\cite{Doerr17} would be a requirement for a new modulator technology to be adopted by the commercial foundries.

The most ‘pure’ electro-optic modulation relies on the Pockels effect, which provides an intrinsically very high E/O BW, even exceeding 100~GHz~\cite{Alloatti14, Weigel18, wangLN18, Burla19, LiMingxiao22}, but these materials face challenges with CMOS integration, and have little or no prior history of integration with CMOS for electronics (compared to Ge and SiN which were already introduced in CMOS electronics). 
LNOI modulators~\cite{Weigel18, wangLN18} provide low IL and have been integrated with sources and PD~\cite{LiMingxiao22}. However, their $\alpha V_{\pi}L$ product needs to be improved further. Lithium, a contaminant in CMOS foundries, restricts FEOL integration. Hybrid integration of etched LNOI modulators~\cite{Wang18} (both MZMs and MRMs) to silicon PICs remains a pragmatic solution. Heterogeneous integration of unetched LN to silicon is achieved with BEOL integration or encapsulation technologies. Avoiding etching also avoids structural defect formation, Nb depletion, and heat and pyroelectric charge build-up issues associated with etched LNOI processes~\cite{Weigel18}. Conversely, improving modal confinement becomes difficult. Since the optical mode is controllably distributed in the unetched LN slab and edge Si rib waveguides, achieving a sharp bending radius for MRMs remains challenging~\cite{Mookherjea23}. The large size of LNOI/Si modulators also prevents adoption in applications requiring many modulators.

Polymer silicon-organic-hybrid (SOH)~\cite{Alloatti14, Eschenbaum22} and plasmonic-organic-hybrid (POH)~\cite{Burla19} require poling and hermetic sealing, creating significant challenges to making stable devices. Their high-temperature reliability and reflow compatibility need to be further demonstrated, although recent results are promising~\cite{Eschenbaum22}. POH modulators, even though they look attractive in the PPA metrics, further suffer from compatibility with CMOS SOI foundries. Good plasmonic metals (Cu, Ag, Au) are also serious contaminants, and need diffusion barrier layers (e.g. TaN) which are optically very lossy.

Polycrystalline layers of other ferroelectric thin-film materials such as BTO show much larger Pockels coefficient (expressed in pm/V) than LNOI~\cite{Czornomaz22} and comparable to polymers~\cite{Xu22}, and recent demonstrations of large E/O BW~\cite{Eltes19, Eltes23} make them promising. Note that a large Pockels coefficient \textit{in} the device is important, which requires a good overlap of the electric modulation field and the propagating optical mode~\cite{Taghavi22, Eschenbaum22}. As part of the direct wafer bonding process, BTO thin films are fabricated first using molecular beam epitaxy deposition on donor wafers, and then directly wafer bonded to interlayer dielectric/SiO$_2$ of the planarized acceptor SOI wafer using intermediate alumina layers as an adhesive. BTO also requires poling to compensate for hysteresis from ferroelectric domain switching (albeit only $\sim$1V DC compared to much larger voltages needed for polymer modulators)~\cite{Eltes19}. Sources of propagation loss include scattering from waveguide sidewall roughness and residual oxygen vacancies in the BTO thin film - areas for further improvement. BTO also has a lower refractive index compared to Si (n$\rm{_{BTO} =}$ 2.38, n$\rm{_{Si} =}$ 3.47 @ 1550 nm) and a significantly large RF relative permittivity that can increase capacitances and velocity mismatch between the optical and electrical fields.

Improving all PPA metrics and HVM suitability is crucial for commercial foundries and LSI applications. However, thanks to the numerous photonic applications, there will always be a need for exceedingly high E/O BW modulators, and several prototyping and R\&D foundries will continue to address the related fabrication challenges. Finally, although $>$100~GHz E/O BW modulators are attractive for both telecom and data center applications, they require electronics capable of driving them at such speeds. Unless $V_\pi$ (or $V\rm{_{pp}}$) is reduced significantly, such electronics will consume a lot of power, regardless of CMOS/BiCMOS/III-V implementation. 

\subsubsection{Phase-Shifters for tuning and switching}\label{PhShift}
Many photonic applications require phase shifters that consume little or no power and have a low $\alpha V_{\pi}L$ for configuration, tuning and switching. For certain applications, these phase shifters should be fast as well, but 10s of GHz E/O BW is not needed. While in many circuits light only passes through one high-speed modulator, it will have to traverse many low-speed phase shifters for tuning and switching, thereby compounding the penalty of power consumption and $\alpha V_{\pi}L$. Metal heaters (or doped waveguides) utilizing the thermo-optic effect are available in all foundry platforms today. They have 1-10$\mu$s response time, and consume considerable power, generating thermal crosstalk, and thus limiting LSI/VLSI scaling. But they do not introduce optical loss, a significant advantage over other alternatives. Improving thermal insulation reduces their power consumption by $>10\times$ at the expense of an even higher response time~\cite{Alqadasi22}. Even $>100\times$ improvement is possible by folding the waveguides to increase interaction with the metal heaters, but that results in an IL~\cite{Lu15}.

The final set of materials and techniques listed in Table~\ref{tab:modPPA} are attractive alternatives to heaters. They include liquid crystals (LC), MEMS/NOEMS, and phase change materials (PCMs). LC on silicon (LCOS) tuning for display applications has been demonstrated at a large scale, and LC has also been the technology of choice for free-space wavelength-selective switches. As phase shifters, they leverage birefringence to demonstrate a strong electro-optic effect. The alignment of the LC molecules can be controlled by applying electrical voltage ($<$1V) without drawing any significant static or dynamic current (nA). Hence, they consume extremely low power but currently suffer from IL~\cite{VanIseghem22}, although very low IL has been demonstrated in visible band recently~\cite{Notaros22}. The liquid integration on chip brings its own set of temperature and packaging challenges, both at the BEOL manufacturing and packaging stages, and requires steps such as etching, inkjet spotting or injection without affecting other devices such as grating couplers, initial LC alignment, and sealing. However, the challenges are surmountable. 
PCM-based non-volatile memory has achieved HVM in the electronics industry~\cite{Optane19} and is being explored for neural network applications \cite{Mukherjee21}. The use of PCM in silicon photonics promises compact tuning capabilities, where the optical phase shift is obtained by tuning the state of the material from amorphous and crystalline. As non-volatile phase shifters, they can sustain their state without any static power consumption. But they suffer from IL and significant dynamic power consumption~\cite{Rios22, Yang23}, rendering them suited to only selected applications where sporadic phase shift is needed. MEMS/NOEMS-based phase shifters are inherently low power~\cite{Feng20}, and have been demonstrated with multiple foundries~\cite{Ramey20, AIM21, Edinger21}. A promising phase shifting mechanism in NOEMS uses the applied voltage to mechanically move the waveguide structure, changing the optical mode field distribution and hence the effective refractive index~\cite{Baghdadi21}. A dual slot structure can be used~\cite{Baghdadi21}, where the dual slots (actuation regions) within a p-i-n junction act as capacitors that get charged or discharged with the applied voltage without drawing significant current, at speeds comparable to metal heaters but without the thermal crosstalk. Compact length, $<$1V drive voltage, low power, and negligible IL~\cite{Baghdadi21} make NOEMS phase shifters an appealing choice for the next-generation phase shifter technology in silicon photonics. Challenges such as hermetic sealing with optical and electrical feedthroughs are solvable~\cite{Midolo18, Jo22}. Finally, materials such as BTO promise high-speed modulation and compact, low-power phase shifting~\cite{Ortmann19}, at the cost of very steep technical and economic integration challenges. 
As phase shifters, their IL must be considerably reduced to compete with other technologies. Integration of emerging materials such as graphene~\cite{Sorianello20} and Indium Tin Oxide (ITO)~\cite{Gui23} into silicon photonics has been demonstrated. Being relatively newer technologies and friendly to CMOS SOI integration, more progress is expected to improve their performance.

\subsection{Laser Integration}\label{Laser}
\begin{figure}[h]%
\centering
\includegraphics[width=0.9\textwidth,trim=2.8cm 4.5cm 3.2cm 2cm, clip]{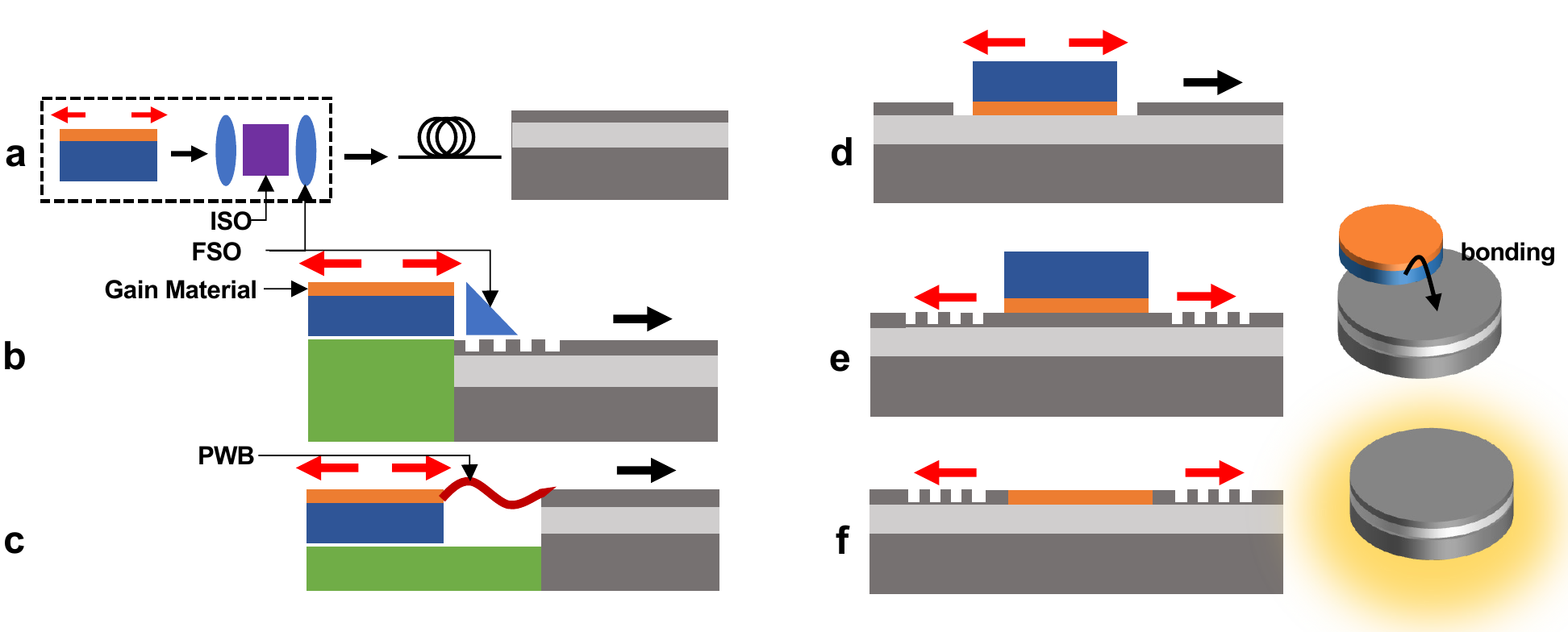}
\caption{Techniques to attach a laser to a silicon PIC. a.  Conventional laser-isolator (ISO)-fiber-PIC with free-space optics (FSO) b. Hybrid 2.5D with FSO. c. Hybrid 2.5D with photonic wire bonding (PWB). d. Hybrid 3D (flip chip or transfer printing). e. Heterogenous (Direct bonding or transfer printing) f. Monolithic (Hetero-epitaxy).
}\label{figlaser}
\end{figure}

Silicon’s indirect bandgap prohibits efficient optical gain that is necessary for a  laser (CW carrier) on the PIC. This deficiency requires alternative materials or methods to introduce light sources on a silicon chip, and the developments over the past decades have led to different solutions (Fig.~\ref{figlaser})~\cite{Reza23}. The conventional technique is to fiber-attach the PIC with a laser and an isolator (Fig.~\ref{figlaser}a). More scalable approaches integrate III-V gain materials with the PIC without fiber. But an isolator is still needed if the laser cannot tolerate reflections. Off-chip isolators perform well but are bulky and increase packaging complexity and cost. Pragmatically, it is often possible to design chips and packages in such a way that back-reflections are not a limiting factor; the high losses in the transmit path provide a barrier between the outside world and any light source. And the cost of compact isolators can be managed when designed into the package.
On-chip reflection-control approaches (Fig.~\ref{figLSI}) that can eliminate the need for bulky isolators include carefully designing the photonic components to reduce reflections below the tolerance threshold of the laser, reducing the reflection sensitivity of the laser by using quantum dot gain regions with low linewidth enhancement factor~\cite{Duan19}, monolithic integration of magneto-optical materials (e.g., Ce:YIG)~\cite{Zhang19}, spatiotemporal modulators~\cite{Doerr14}, or active reflection cancellation circuits~\cite{Shoman21}. A generalized, low-cost, scalable, on-chip, low-loss, low-power, and compact solution robust to near- (coherent) and far-end (incoherent) modulated multi-wavelength reflections remains a research problem.

A pragmatic solution for laser integration is hybrid integration, where multiple chips from different material technologies are co-packaged together. For example, (sub)-mm DFB lasers, manufactured by the millions for datacom applications at low cost and high yield and pre-tested, can be co-packaged with a silicon photonic chip or even with a wafer. A 2.5D integration technology that has been commercially successful involves packaging known-good lasers with the silicon photonic die using epoxy, ball-lens, and isolator (Fig.~\ref{figlaser}b)~\cite{Luxtera18}. Other 2.5D techniques include using butt coupling ~\cite{Jin21} or photonic wire bonding ~\cite{Billah18} to enable relaxed alignment tolerances (Fig.~\ref{figlaser}c). These 2.5D techniques are adequate for several bespoke silicon photonic applications today. Hybrid 3D integration technologies (flip-chip or micro transfer-printing) promise to further shrink the assembly size at the cost of using the PIC area (Fig.~\ref{figlaser}d)~\cite{Song16, Guan18, Wang18, Zhang18}, but require high-accuracy placement and bonding.

\begin{figure}[h]%
\centering
\includegraphics[width=0.99\textwidth,trim=0.02cm 0.02cm 0.02cm 0.02cm, clip]{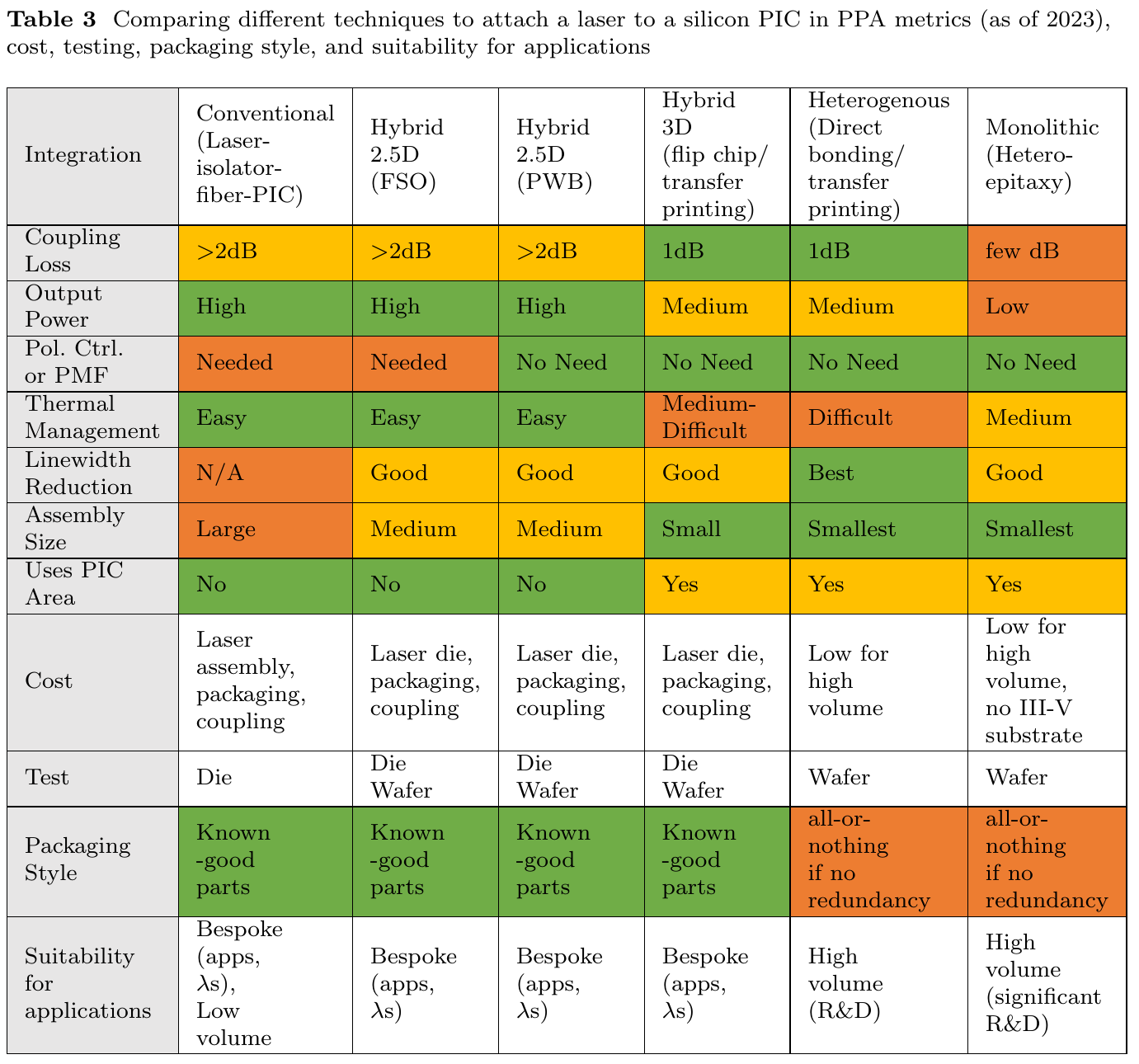}
\label{tab:LasertoPICPPA}
\end{figure}

Table~\ref{tab:LasertoPICPPA} summarizes the PPA and other metrics for various integration schemes. The wall-plug efficiency (WPE) of most of the C/L/O band lasers is only about 10\%, a metric that needs more focussed research for improvement. With similar WPE, the coupling loss between the laser into the PIC can be instead taken as a proxy for power consumption. In the 2.5D hybrid integration, a separate laser provides the flexibility of choosing the one with the needed optical power, and thermal management is easy. For higher power handling, SiN can be used on the PIC. There are means to improve the linewidth of the laser from the typical linewidth of a DFB and also cancel the reflections to improve isolation~\cite{Shoman21}. 3D techniques with high-Q Si or SiN external cavity support linewidth reduction down to 1 Hz and lower~\cite{Jin21, Li21, GuoJ22}, more than sufficient for applications such as coherent communication~\cite{Guan18} and automotive LIDAR. Hybrid integration also allows multiple wavelengths~\cite{Wang18}. Still, the benefits of hybrid integration vis-a-vis scaling towards $>8 \lambda$ WDM LSI PICs needing multiple lasers, gain elements, etc., remain to be thoroughly demonstrated. 

Another commercially successful technique in HVM ($>$ million/year)~\cite{Fang06} has been heterogeneous integration, where multiple materials or epitaxial stacks are processed together into one silicon chip at wafer scale. Again, various strategies have been adopted~\cite{Juniper13, HP16}. They include bonding III-V chips to Si with coarse alignment followed by post-processing the Si wafer to make quantum-well (QW) lasers (Fig.~\ref{figlaser}e)~\cite{Fang06, Alduino4x10, Jones19, Fathololoumi22}. Thermal isolation of the gain medium by the buried oxide (BOX) and the mismatched CTE must be carefully addressed for high-temperature operation, efficiency, and reliability. Placing redundant lasers helps improve the failure-in-time (FIT) rates~\cite{Fathololoumi22}. Benefits of the heterogeneous approach include sub-dB coupling loss and a mechanism to leverage the low-loss external cavity in silicon to significantly reduce the laser linewidth using self-injection locking~\cite{Kondratiev22}. 

Another longer term approach, desirable for quantum-dot (QD) lasers, is to directly grow epitaxial gain material on the Si wafer~\cite{Liu15}. Due to its lower linewidth enhancement factor, $\alpha\rm{_H}$, QD lasers enable lower linewidth and reduced sensitivity to reflections~\cite{Norman18, Duan19}. They also have lower threshold current density. Monolithic integration using hetero-epitaxial growth (Fig.~\ref{figlaser}f), where the III-V substrate is not even needed, remains the end goal, with several recent progress and more to come~\cite{Margalit21}. 

Multiple silicon photonic foundries are developing hybrid or heterogeneous laser solutions. For scalability purposes, the foundries will likely favor a technology that lends itself to multi-wavelength support, which is crucial for several LSI applications. It is likely that scale-out will be supported first by bonding multiple single-wavelength lasers~\cite{Fathololoumi22}. Comb lasers~\cite{Chang22} such as passive semiconductor mode-locked lasers (SMLLs)~\cite{Liu19} are being actively pursued by various research groups. DFB arrays ensure large output optical power in each wavelength, whereas, in the SMLLs, the power is split between the wavelengths. The presence of the saturable absorber further lowers the total (and hence per wavelength) output power of SMLLs.  
However, an SMLL is significantly smaller than a DFB array. The linewidth of passive SMLLs~\cite{Liu19} is usually lower than DFB arrays~\cite{Fathololoumi22}. More R\&D is expected for SMLLs to demonstrate higher power, reliability, and lifetime in the next decade. Such requirements for DWDM applications are even more stringent, and any temperature drift creates inter-channel crosstalk~\cite{Chen15, Jayatilleka16}.

\subsection{Avalanche Photodetectors}\label{APDs}
Most of the silicon photonic applications are constrained by limited output power and WPE of the laser, and the high IL in the circuits. An alternative is to improve the signal-to-noise ratio (SNR) at the detection stage (Fig.~\ref{figLSI}). Low-voltage APDs which have large -3dB O/E BW, high overall responsivity~\cite{Chowdhury22}, and simultaneously low noise will be beneficial for receiver SNR improvement~\cite{Benedikovic22, Kang08}. It is important to note that the overall responsivity (in A/W) and low noise is crucial. A large multiplication gain for an APD which has a poor intrinsic responsivity does not lead to a superior performance. Although relatively easier to achieve in Si APDs at 850~nm ~\cite{Nayak19}, simultaneous (at the same bias voltage) optimization of gain-BW-noise has remained challenging for low-voltage Si/Ge APDs~\cite{Wang22} or Si resonant APDs~\cite{Sakib20, Peng23} in C/L/O bands. In comparison to Ge PDs, APDs generally have inferior BW, linearity and power handling, which limits their use in various applications. APDs also need to be biased optimally and stabilized for temperature and voltage drift, but that is less challenging~\cite{Nayak19} than what has already been demonstrated for microring circuits~\cite{Sun19}.  

\subsection{Delay}\label{Delay}
Several silicon photonic applications require hundreds of picoseconds to nanoseconds of delay. Examples include microwave photonics, optical phase-locked loops (OPLLs), frequency discriminators (Fig.~\ref{figLSI}), laser linewidth reduction circuits, OPAs, optical coherence tomography (OCT), and gyroscopes. Many of these applications also require tunability in 10s of picoseconds and broadband operation~\cite{Ji19}. Realizing such a delay in silicon photonics with low-loss and low-area has been very challenging~\cite{Hong22}. Resonant devices provide a narrowband delay. Si or SiN delay lines are difficult to tune and require narrow bends leading to significant scattering and radiation losses. Shallow etched ridge waveguides or ultrathin waveguides break compatibility with the 220-nm processes. Modifying the fabrication process without sacrificing the performance of other photonic components remains challenging~\cite{Xiang23}.

\section{Silicon Photonics: Systems Perspective}\label{SysPers}

\subsection{Photonics \& Electronics Interplay}\label{EPICPlay}
Silicon PICs almost always exist in conjunction with electronic ICs (EICs). When we look at systems based on photonic chips, the landscape today is almost 100\% dominated by data communication, and we expect this to continue for the near future. In this context, EICs serve two purposes (Fig.~\ref{figLSI}): (1) Enable \textbf{E}/O and O/\textbf{E} conversions of the end-to-end data. (2) Bias, control and compensate for temperature and fabrication variations. Thus, photonics serve electronics by providing the data links, and electronics serve photonics by providing control and readout and digital signal processing (DSP). A major difference between photonics and electronics is that photons don’t interact and thus are excellent for transmission of information, whereas electrons interact and repel each other and thus make good switches and computing elements. Each silicon photonic switch therefore requires a corresponding electronic switch. On the whole, the number of transistors in the EIC that must accompany an LSI PIC are orders of magnitude larger than the number of components in the PIC. Here lies a natural interplay, since transistors consume much lower power in (1) switching, (2) providing gain (both linear and limiting), and (3) offering high precision, while being orders of magnitude smaller than the photonic components~\cite{Shekhar21}. On the other hand, the photonic components (1) enable lower frequency-dependent loss when moving data over a longer distance compared to copper, (2) may provide lower latency through asynchronous and repeaterless data movement, and (3) ease parallelism of very high-speed data on an optical waveguide (through WDM). When the data is already in the optical domain, photonic signal switching or processing can become attractive. The former is a widely deployed technology, while the latter has yet to make the leap from research to product to replace DSP functionality. Thus, it is good to be cognizant of the respective virtues of the PIC and EIC technologies. For example, the E/O and O/E overhead of processing electronic data in the photonic domain must be carefully analyzed. Conversely, silicon photonics provides opportunities to shrink large optical systems, and bring new applications (such as in sensing and imaging) to reality, which electronics cannot enable by itself. Finally, silicon photonics operates on a carrier wave of hundreds of THz, while silicon electronics is limited to sub-THz. Such differing attributes open attractive co-design opportunities, such as designing electronic clocks with ultra-low phase noise~\cite{Li13}. 

\subsection{Photonics \& Electronics Ecosystem}\label{EPICEco}
It is insightful to look at the electronics industry ecosystem briefly. Moore's law demonstrates that the cost per component goes down with every generation of CMOS technology reducing the critical dimensions of the transistors. This scaling is enabled by an exponential increase over time in the economic scale of the semiconductor industry, which allows the industry to pay for ever more expensive foundries and process development. Foundries enable many users to access these advanced processes, without each needing to pay to develop the process on their own. At the most extreme, the MPW (multi-project wafer) runs that the foundries host allow multiple users to share the costs of a single wafer run to develop products cost-efficiently.

As processes mature, yields go up, and costs come down. The foundries and third-party intellectual property (IP) providers enable a process design kit (PDK) and design IP libraries, allowing the customers to build incredibly complex electronic circuits and get them right the first time. By relying on both proven devices and proven circuit-level IP, the designers can focus on system-on-chip (SoC) integration without ever touching the transistor level in several cases.

Once the chips are fabricated, there is a rich ecosystem of test houses, packaging service providers, and so forth. Electrical wirebonding (Fig.~\ref{figEPIC}a) and flip-chip bonding (with C4 bumps and microbumps, Fig.~\ref{figEPIC}b) are reliable and popular means of packaging, with the latter providing more bumps instead of just peripheral connections. More advanced packaging techniques (see Fig.~\ref{figEPIC}) such as through-silicon via (TSV), TSV-less interposers, and heterogeneous integration are used to improve signal integrity, power and thermal distribution, and die yield by breaking complex and large SoCs into smaller chiplets~\cite{Lau22}. Because the FPGAs, GPUs, and CPUs are produced in HVM, the overall cost still goes down despite the complex packaging techniques. Nevertheless, judicious packaging decisions are made to avoid unnecessary complexity; generally, the simplest package is best, and advanced packaging techniques (chip on wafer, chip stacking, etc.) tend to be introduced only when no other alternative is feasible.  

The photonics industry has several similarities but also many stark differences. Just like in the electronics industry, increasing the number of photonic components is not always about reducing cost, but is often about providing new functionality, improved performance, or reduced area per component. MPW runs are now available at many foundries, although mature PDKs, and abstraction languages are still in very early stages. Third-party IP support is mostly non-existent thus far. Companies wall off the most advanced PIC processes to protect their investment and IP (reminiscent of the early decades of the CMOS industry, acting as virtual integrated device manufacturers (IDMs), maintaining differentiation at the process and PDK level. Meanwhile, academic research mainly focuses on improving the devices. 

Photonic foundries face a significant dilemma: Their customers often demand that they customize their processes, which involves a great deal of R\&D expense, and endangers the reliability and yield of the final wafers. Driving customers into a standard process is the solution for this, but in order to do that, the customers need to see significant value in stability and in a settled PDK and IP ecosystem; only a few designers see the world this way, because so many of the members of the design community today were trained as device people, rather than SoC designers. Changing process parameters often seems to such designers to be the easiest way to generate performance differentiation, but the downstream costs for such changes can be very high from a reliability and process maintenance perspective. As more designers who are used to the idea of settled PDKs graduate and come into the field, disruptive process changes will slowly become less and less common; the foundries will also likely grow ever more resistant to process changes from customers that are not justified by substantial purchase commitments.

The overall yield for silicon photonics products is still lower than their CMOS electronic counterparts. Additional factors at the process, design, and packaging level account for the difference: fabrication~\cite{Bogaerts18, Xing23} and thermal sensitivity, lack of robust PDK components and variation-and-mismatch aware models~\cite{Bogaerts18, Xing23}, design flow methodologies still missing hierarchical simulations, schematic driven layout and layout-versus-schematic verification~\cite{Bogaerts18}, custom process modifications for specific components, challenges with epitaxial growth, Ge integration for photodetection, integration of laser (whether at the die or package level), laser FIT, and fiber connectivity. Only a handful of HVM silicon photonics products are shipping today, requiring the fab to timeshare the production with other processes, and adding another source of yield impact.

\subsection{Photonics \& Electronics Co-integration}\label{EPIC_coint}

\begin{figure}[h]%
\centering
\includegraphics[width=0.9\textwidth,trim=4cm 2cm 4cm 2cm, clip]{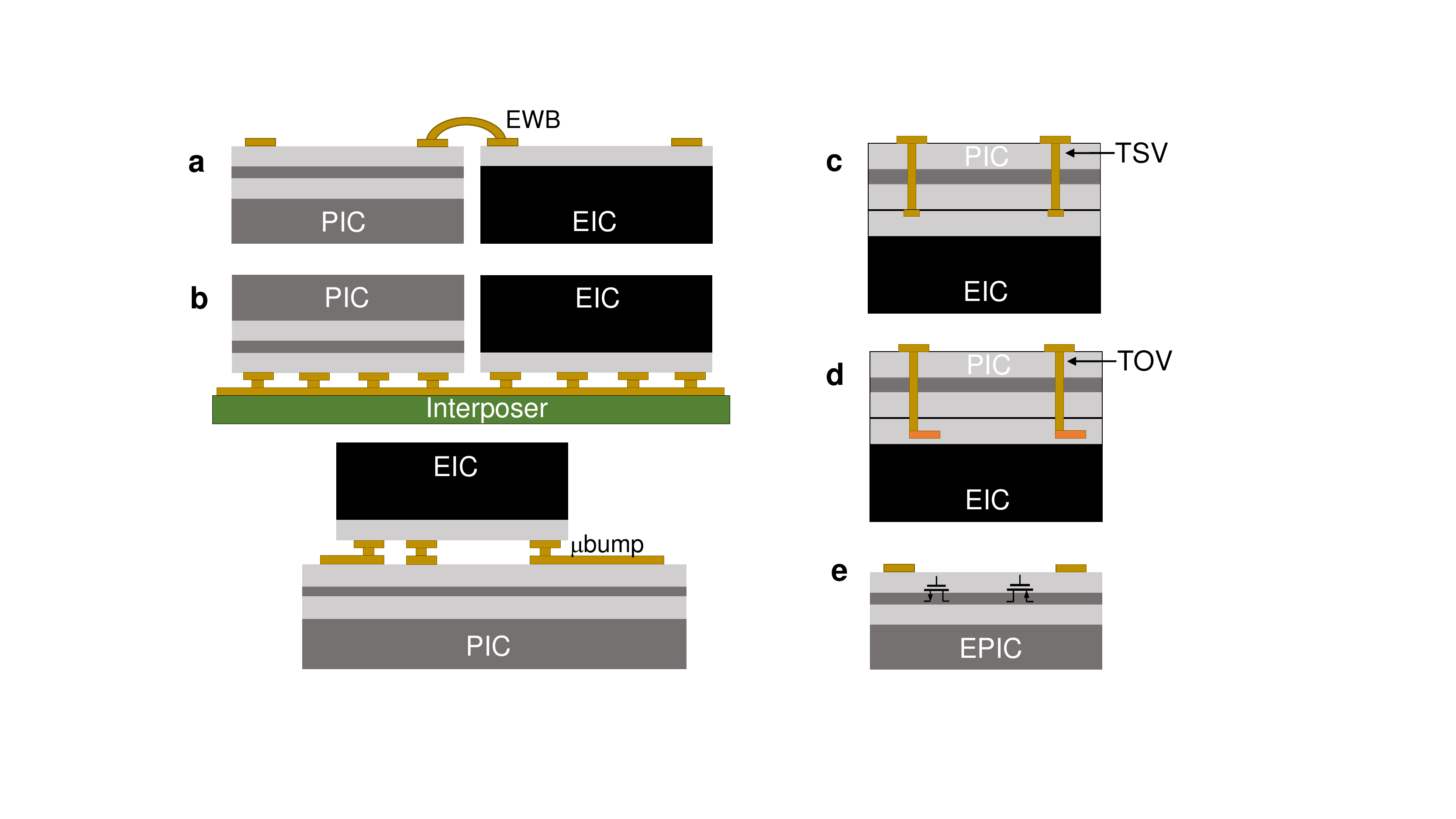}
\caption{Comparing different techniques to attach a PIC to an electronic IC (EIC): a. Electrical wire bonding (EWB) side-by-side b.  2.5D flip-chipped side-by-side or stacked c. Hybrid 3D TSV (Through-Silicon Via)  d. Heterogenous 3D with TOV (Through-Oxide Via)  e. Monolithic electronic photonic IC (EPIC).
}\label{figEPIC}
\end{figure}

\begin{figure}[h]%
\centering
\includegraphics[width=0.99\textwidth,trim=0.02cm 0.02cm 0.02cm 0.02cm, clip]{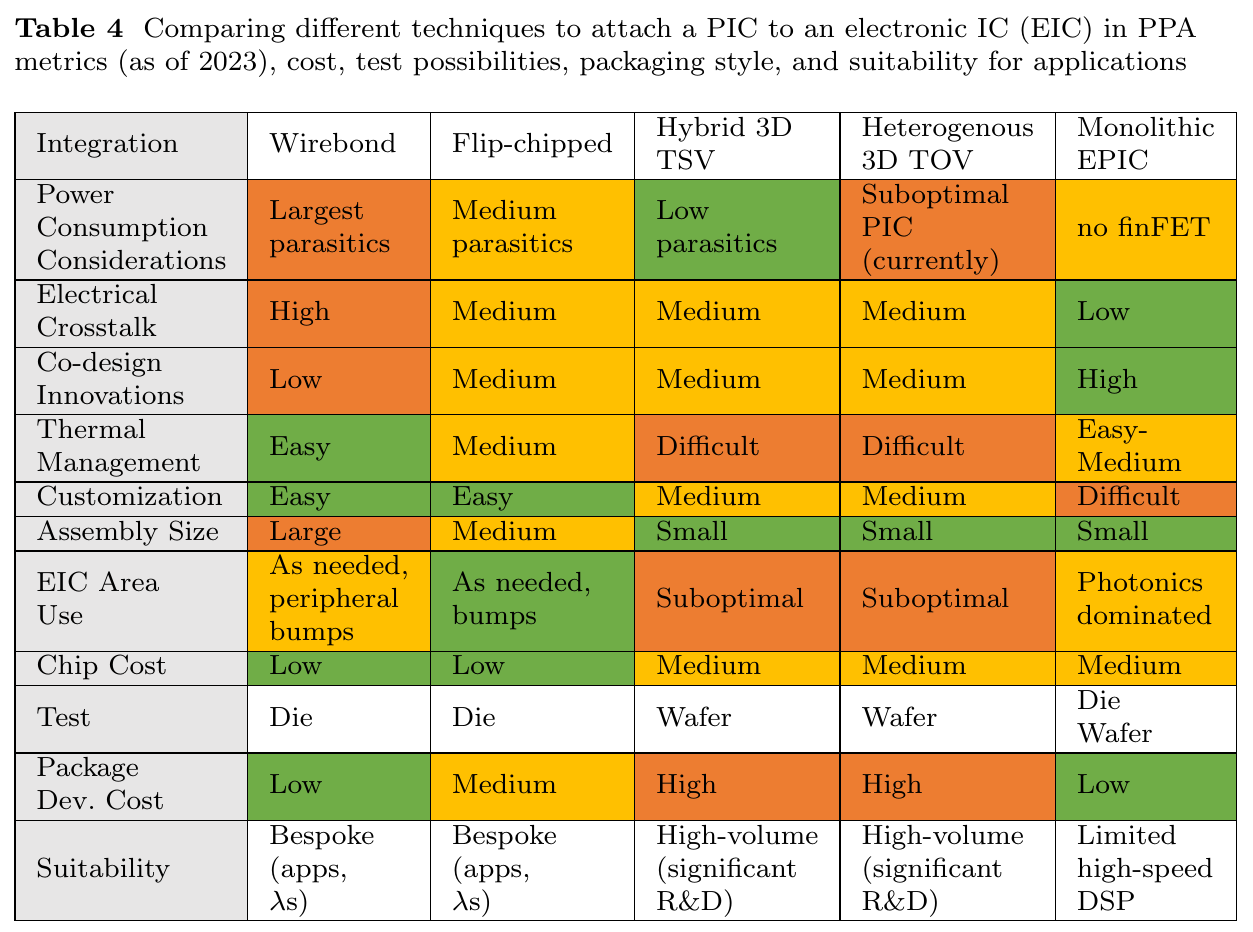}
\label{tab:EtoPICPPA}
\end{figure}

The option to integrate the PIC with the EIC has been around since the first commercially successful silicon photonic product~\cite{Margalit21, Stojanovic18,Bogaerts18}. Developing a monolithic EPIC process (Fig.~\ref{figEPIC}e), starting with a CMOS (or BiCMOS) SOI process and optimizing it for photonic applications, has been demonstrated several times~\cite{Gunn06, Zimmermann15, Rakowski20} successfully. From the perspective of commercialization and time-to-market, a monolithic EPIC often `seems to' be the superior technology of choice (Table~\ref{tab:EtoPICPPA}). High-speed circuits such as drivers and TIAs can be colocated next to modulators and PDs, reducing parasitics and power consumption~\cite{Giewont19}. Controllers (thermal, wavelength) can be designed and placed next to the photonic components without needing dedicated pads. For LSI applications, a monolithic EPIC can simplify packaging complexity significantly. However, when the die area is dominated by photonics, photonic components being orders of magnitude larger than their electronic counterparts~\cite{Shekhar21}, the overall die cost can increase significantly without arguably making full use of CMOS devices. This analysis has to be done case-by-case for individual products.

In principle, microring-based circuits appear to be very appealing for monolithic EPIC processes until the next-generation modulator with a superior FoM$\rm{_{efficiency}}$ is developed (Section~\ref{Mod}). But to conclude whether they make sense in a given, specific application, a complete systems analysis is necessary; microrings come with considerable control overhead and performance tradeoffs, especially at very high speeds. If the application requires high-speed ADC/DAC and especially DSP (Fig.~\ref{figLSI}), another finFET EIC must also be added to save power consumption, as the fastest monolithic EPIC process today in 45-nm CMOS SOI is still several generations slower (in fanout delay) than finFET processes. Integrating photonics directly onto CMOS wafers below the 45 nm node is unlikely to occur in the next few years; doing so does not make economic or technical sense in a world where chip-on-wafer bonding between PICs and scaled microelectronics is comparatively straightforward.

Other possibilities for EPICs have also been explored. Adding photonics to an older generation CMOS process leads to high-power and slower drivers and TIAs, leading to worse transceiver designs and rendering them unattractive to the biggest customers of silicon photonics - datacom and telecom. Nevertheless, such a process is appealing to university researchers as it opens up opportunities to co-design and innovate new EPIC circuits~\cite{Idjadi17, Moazeni17} at low cost and packaging effort. On the other hand, multiple efforts are underway to integrate transistors onto the same wafers as silicon photonic devices~\cite{Zanetto22}. However, doing so has thus far involved unacceptable compromises to the performance of the bipolar electronics. 

Most of the silicon photonic transceivers in HVM today are based on a 2.5D integration approach, where the PIC and EIC(s) are designed, sized, optimized, tested in their best respective processes, and then flip-chipped to an interposer substrate~\cite{DoerrOFC14, Luxtera16, Jones19, Ahmed19, Ahmed23} (Fig.~\ref{figEPIC}b, Table~\ref{tab:EtoPICPPA}). The EIC process can be chosen from one of the many CMOS/SiGe foundries. Multiple EIC chips can also be flip-chipped, such as (1) a SiGe chip or scaled-CMOS chip with a reasonably large breakdown voltage to permit high-swing drivers and a reasonable switching speed to support the RF speed requirements, and (2) an advanced FinFET chip for DSP/ADC/DAC~\cite{Rakowski18}. An EIC process with faster transistors may even compensate for the parasitic capacitance due to additional pad, ESD, and routing (compared to a monolithic EPIC solution). For LSI applications where most PIC components require electronics at a relatively low speed (such as LIDAR), flip-chip solutions seem reasonable~\cite{Poulton22}. However, for LSI applications that need many high-speed drive/readout lines, a flip-chip solution means many RF traces on the interposer, leading to complexity and crosstalk considerations. In either case, the size of the PIC is increased due to the necessity of many I/O bumps, though with microbumping and copper pillar technologies to realize a stacked flip-chipped 2.5D package~\cite{Boeuf13,Dobbelaer17} (Fig.~\ref{figEPIC}b, bottom), these increases are often commercially negligible. The parasitics and interconnects are also reduced compared to their side-by-side counterparts.
A hybrid 3D integration can be considered in some cases, where the EIC is flip-chipped on the (larger) PIC chip and uses advanced techniques such as TSVs or through-oxide vias (TOVs) (Fig.~\ref{figEPIC}c, Table~\ref{tab:EtoPICPPA}). The RF lines still need to be routed from the small EIC to several places on the PIC, which remains challenging. A WoW heterogeneous 3D integration is also being researched where the photonics wafer is flipped and vertically attached with the CMOS wafer through oxide-bonding, the silicon handle on the photonics wafer is removed, and TOVs are formed at the wafers’ interface~\cite{Uzoh16,Kim19}; further improvements are expected for the performance of photonic components in such an integration technology (Fig.~\ref{figEPIC}d, Table~\ref{tab:EtoPICPPA}). One possibility is to use multiple EICs 3D integrated on the PIC. 

Overall, the application, performance specifications and the volume of shipments (affecting the cost) will decide whether a more expensive monolithic EPIC with simpler packaging, a multi-chip 2.5D integration with more complex packaging, or a 3D integration with more complex processing/packaging is the right choice (Table~\ref{tab:EtoPICPPA}). We expect that all of these scenarios will co-exist, just like in the electronics ecosystem. 

\section{Silicon Photonics: Applications Perspective}\label{App}

\begin{figure}[h]%
\centering
\includegraphics[width=0.99\textwidth,trim=0.02cm 0.02cm 0.02cm 0.02cm, clip]{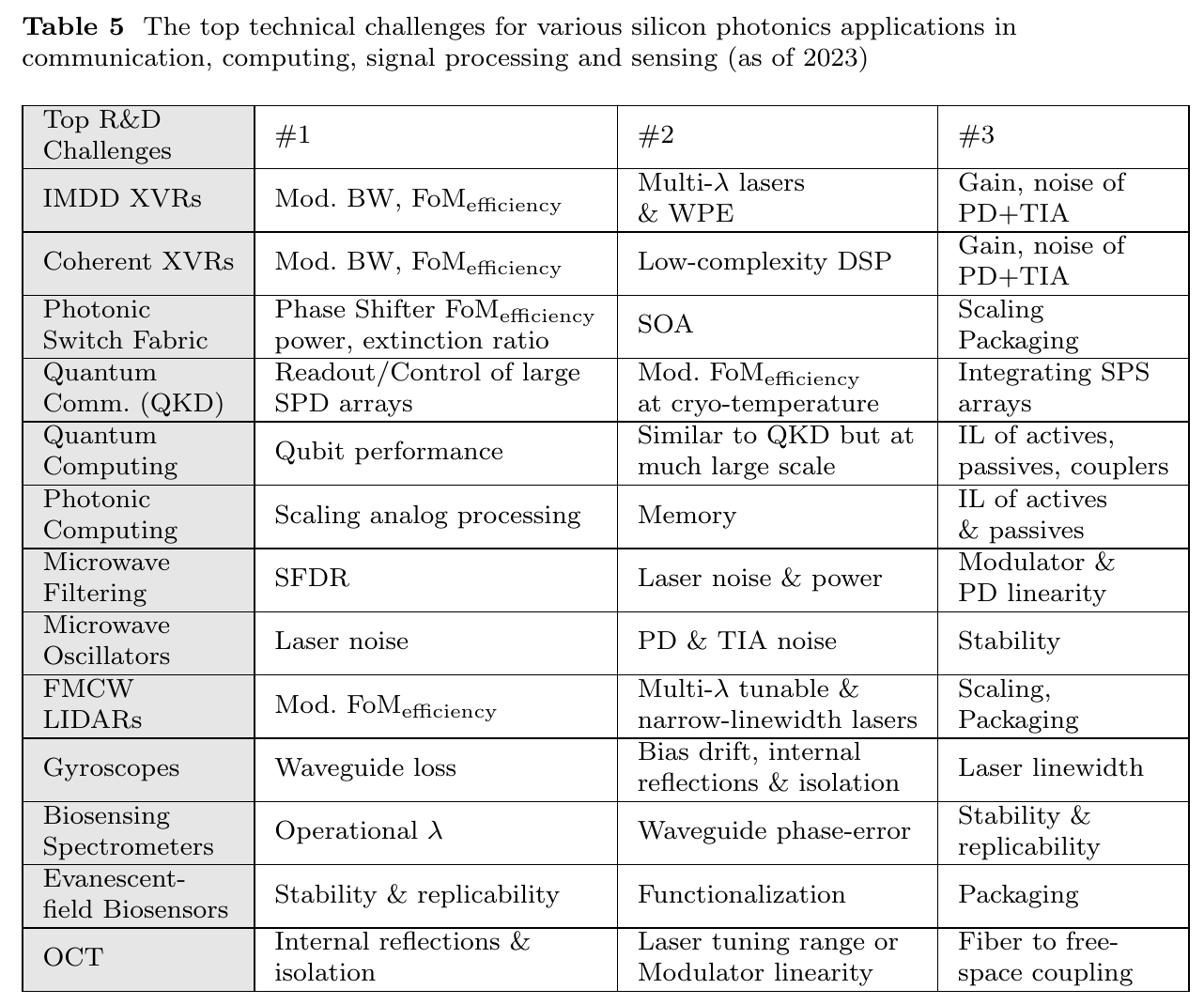}
\label{tab:Apps}
\end{figure}

In this section, we describe the top technical impediments to the success of various silicon photonics applications (Table~\ref{tab:Apps}), connecting them to some of the challenges and opportunities discussed in previous sections. We limit the impediments to PIC/EIC technology only, excluding economic, regulatory, market, and other factors such as chemistry, biomarkers, quantum advantage, etc. We also do not delve into the benefits of silicon photonics for these applications since most of the previous works describe them in detail. 
 
For IMDD transceivers (XVRs) to further improve their energy efficiency (pJ/b) and scale to higher data rates, the modulator FoM$\rm{_{efficiency}}$ needs further reduction, and the $-$3dB E/O BW needs to be improved towards 100~GHz. Improving the WPE of lasers is essential for most applications but especially crucial for communication and computing applications. Efficient multi-wavelength light sources are also needed with adequately large power in each wavelength. Low-noise, large gain-bandwidth APDs in O/L/C bands could provide an SNR improvement without significant power consumption penalty, but historically their bandwidth, linearity, noise and power handling characteristics have prevented their use at the highest bandwidths. Finally, amplifying PD signals using high-gain, low-noise TIAs remains a crucial challenge. Several equalization-based techniques have been recently demonstrated to limit the noise using low-BW TIAs~\cite{Ahmed21}, but most operate on the assumption that the receiver clock is available.  

For coherent transceivers to be competitive inside data centers, additional challenges (vs. IMDD) must be solved. Linearity requirements for the TIAs and drivers are more stringent~\cite{Ahmed20, Ahmed23}, and the reliance on power-hungry DSP needs to be reduced as much as possible. One strategy being explored by researchers is moving some signal processing tasks into the optical domain~\cite{MorsyOsman18, Hirokawa21} leveraging integrated photonics and analog electronic circuits. The latter requires significant electronic-photonic co-design effort, opening up several opportunities for CMOS designers to leverage the expertise from mixed-signal and RF ICs. 

High throughput network switches for short-reach to long-haul markets require the phase shifters to have excellent FoM$\rm{_{efficiency}}$ to enable large fabrics. The switching must incur low power consumption, low loss and demonstrate a large extinction ratio. For applications that permit slower switching speeds, insulated metal heaters in interferometric switches are currently the popular implementation choice~\cite{Lee19}, but technologies such as MEMS/NOEMS look promising~\cite{Seok19}. Long-term reliability and demonstration in large-scale fabrics co-integrated with electronics and packaged with optical I/Os are needed. Polarization diversity and wavelength considerations further complicate the scaling and packaging considerations. Applications requiring fast switching are even more challenging since high-speed modulators with comparatively inferior FoM$\rm{_{efficiency}}$ further deteriorate IL and extinction ratio. Regardless of the switching speed requirements, the inherent losses in large switch fabrics require optical amplification, necessitating the integration of SOAs, ideally uncooled, for energy efficiency considerations.

Practical quantum communication and computing applications require LSI-VLSI photonic components with advanced CMOS controllers. For chip-scale discrete-variable quantum key distribution (QKD), the foremost requirements are the cryo-compatible photonic/electronic readout and control of superconducting nanowire single-photon detector (SPD) arrays; developing low-loss, low-power cryo-modulators and cryo-compatible WDM mux/demux; and integrating single-photon source (SPS) arrays at the transmitter in a low-noise, low-crosstalk chip-scale photonic-electronic solution. Superconducting nanowire SPDs operate at telecom wavelengths, facilitating the use of existing optical fibers as a quantum channel. Besides massive parallelization, reducing the loss in the receiver and improving the SPD performance will help increase the transmission rate~\cite{Beutel22}. For quantum computing applications, the challenges are similar, but require much larger scalability of qubit control/readout, including the photonics and low-latency control electronics~\cite{Vigliar21}. The quality of qubits is, of course, paramount. Scalability of control/readout degrades with IL - every photon lost degrades the capability of the quantum system in an exponential way. Ultra-low-loss couplers are therefore needed to connect to the PIC.

Photonic computing involves analog computation and processing of information within the photonic domain~\cite{Bhavin21, Bandyopadhyay22}. This requires handling multi-level signaling~\cite{Guo22} and increasing the precision of weight control~\cite{Tait18} to ensure a high SNR. Such improvements are crucial to achieve accuracy comparable to the incumbent CMOS EIC compute engines~\cite{Alqadasi22}. Another challenge is access to high-speed memory to prevent a memory bottleneck, especially for activations and tasks that are not weight-stationary. Photonic computing uses high parallelism, so it is essential to reduce the IL of passive and active devices (modulators, phase shifters) and boost the output power of multi-wavelength lasers to accommodate larger network sizes. Additionally, for neural networks, efficiently implementing programmable nonlinearities stands out as a significant hurdle~\cite{Bhavin21}.  

For automobile driving, silicon photonics LIDARs are positioning themselves as a solid-state challenger to Time-of-Flight (ToF) LIDARs utilizing mechanical or MEMS-based scanning. LIDARs consist of two subsystems - ranging and beam steering, both of which can use silicon photonics. ToF and frequency-modulated CW (FMCW) are ranging techniques. FMCW provides the benefits of (1) coherently detecting signals down to a few photons, (2) robustness to interference from ambient sources, and (3) simultaneous distance and velocity measurement. All of the necessary components for coherent detection can be integrated on a single chip. For beamsteering, two integrated possibilities exist: (1) Optical phase arrays (OPAs), based on continuous tunable phase shifters and gratings~\cite{Poulton22}. Bulk optics solutions, such as spinning mirrors and oscillating mirrors, have the advantage of being cheap, mature, and simple; displacing such solutions with an on-chip OPA will be a significant challenge. 
For an OPA to emit a single beam, the grating  antennas need to be spaced less than half a wavelength (in free space) - a challenging proposition for 2D beam steering on a silicon chip. Therefore, silicon photonic OPAs typically have gratings arranged for beam steering in 1D and the wavelength of the laser is swept to steer the beam in the other direction. (2) Focal plane arrays (FPAs) based on on-chip switch networks and grating couplers~\cite{Rogers21}. These include 2D FPAs, utilizing MEMS switches~\cite{Zhang21, Zhang22}, or 1D FPA with wavelength steering. Regardless of the solution, low-power (10s of nW) and improved FoM$\rm{_{efficiency}}$ phase shifters are important, and necessary for beam steering. Improved lasers are the next challenge. For 1D OPAs or FPAs, multi-wavelength lasers can relax wavelength tuning~\cite{Riemensberger20}. For FMCW demodulation, narrow linewidth ($<$100~kHz) continuously tunable lasers (preferably without mode hopping) are crucial. Scaling and packaging comprise the third challenge. Scaling the photonics and electronics for many emitters and phase shifters, and integrating a considerable delay~\cite{Ji19,Hong22,Xiang23} (Fig.~\ref{figLSI}) for the laser and a complex DSP, are both necessary. 

Applications in microwave photonics, such as filters and low-phase noise oscillators, have different challenges than most other applications discussed so far. The spurious-free dynamic range (SFDR) specification for microwave filtering is quite challenging to realize in the current generation of silicon photonics. Strict linearity of the modulator and PD is needed, and at the same time, several sources of noise (laser, PD, TIA) have to be minimized~\cite{Liu18}. The goal to get an RF net gain further complicates the design. Implementing laser-assisted microwave oscillators through silicon photonics with phase noise superior to CMOS-only counterparts also requires minimizing the laser noise, PD, and TIA noise~\cite{Xiang23}. Excellent short-term and long-term stability is also needed.  

Silicon photonics promises low-cost and compact gyroscopes~\cite{Liang17}. But to compete with their fiber-optic-based counterparts in performance, gyroscopes in silicon photonics leveraging the Sagnac effect must demonstrate ultra-low loss in waveguides (mimicking an optical fiber), reduction in bias drift due to reflections under extreme conditions of vibration and changing temperature (to not mask the Sagnac phase shift), and low noise for high sensitivity. Engineering of SiN waveguides has reduced the loss to 0.5dB/m~\cite{Xiang23}, and further improvements are needed. Back reflections must be eliminated both on-chip and off-chip, for which on-chip isolators, reflection cancellation circuits, or self-injection locking~\cite{Kondratiev22} must be reliably implemented. Since a gyroscope neither requires an LSI implementation nor high-speed modulators, it is a promising bespoke application that can be successfully realized if the challenges mentioned above are solved. Robustness to vibration also mandates a heterogenous implementation whose challenges must be solved for HVM.  

Silicon photonic spectrometers for biosensing applications often require an operational wavelength incompatible with the C/L/O bands~\cite{Li22}. This becomes the most significant bottleneck, since new waveguides (vs. the standard 220 nm) and other photonic components have to be designed, tested, and characterized~\cite{Zilkie19}. The lasers are also challenging, and the need for broad wavelength tuning or multi-wavelength lasers at these non-standard wavelengths poses severe difficulty. Finally, the stability and replicability of the measurements are crucial for biosensing applications, and the performance of the PIC and laser has to be maintained despite the environmental drift.  

The stability and replicability requirements for evanescent-field biosensors are even more stringent, since an invasive measurement of blood or other bodily fluid increases the user's expectation for trust. The system limit of detection not only depends on the resonator's response to temperature, laser noise, PD and TIA noise, but also on the noise induced by the fluidic flow, mechanical vibration and the biological noise. After the oxide-open step, functionalization of the resonator surface significantly depends on waveguide design and affinity~\cite{Puumala23}. The packaging and integration of the biosensor lead to the next set of challenges. Benchtop machines use an expensive tunable laser and nanopositioners but a simple passive PIC for biosensing~\cite{iqbal2010}. On the other hand, point-of-care devices must be compact, inexpensive, and require operation with a low-cost tunable laser or a fixed-wavelength laser integrated with the rest of the PIC, EIC, and fluidics~\cite{Adamopoulos21, Chrostowski21}. 

Current silicon photonics swept-source OCT prototypes for retinal imaging suffer from poor sensitivity~\cite{Rank21}. First, they operate at either O or C band, while ophthalmology OCT is preferred at 1050~nm for deeper penetration in the tissue. Moving to 1050~nm would require SiN-based PICs and a tunable laser source at that wavelength. Minimizing internal reflections and improving the isolation would enhance the sensitivity. The next impediment is the limited tuning range and sweep rate of the laser source, degrading the image acquisition rate. Finally, the laser power cannot be too high because of laser safety limitations. This, in turn, requires a nearly lossless connection between the PIC and the imaging optics. 

\section{Summary and Conclusion}\label{Summ}
We have made big leaps in silicon photonics - from building the first high-confinement waveguides and the very first modulators only a couple of decades ago - to a technology that has strategically leveraged materials, integration and packaging techniques from the CMOS industry to become the dominant technology in the transceiver space. At the same time, silicon photonics is still very much a technology in development, and a gamut of possibilities, only some of which are described in this article, signify the prospects that lie ahead. Some clear winners will emerge in the next decade and consolidation will happen. Still, the diversity of applications will ensure ample opportunities for the technology to both scale up and spread wide. 

We believe that, in the next decade, we will see the likely realization of the following milestones:
\begin{itemize}
\item Hybrid, heterogeneous and monolithic integration will provide the lasers, phase shifters, modulators, and electronics for LSI and even VLSI implementations with the requisite density, configurability and programmability. Each of these integration techniques has its merits, and will likely coexist for the foreseeable future.
\item Integrated lasers and SOAs on silicon photonics will really take off. Most foundries will provide integrated lasers, with the WPE exceeding 20\%. Both multi-wavelength lasers and tunable lasers will be supported.
\item Silicon photonics will finally diversify beyond pluggable transceivers to other successful commercial products, finding wide adoption for CPO and xPU applications. Complex systems built using interposers and chiplet architectures will adopt photonics for interconnects. Coherent photonics will flourish further - inside communication (even inside data centers) and sensing (FMCW LIDAR, biosensing).
\item The design, modeling, simulation, fabless manufacturing, packaging, and test ecosystem will start to mature, bringing a new cohort of engineers and increased access. A shorter fabrication turnaround will further expedite the R\&D.
\item Plasma-dispersion-based modulators will continue to serve adequately for many WDM communication applications, but at the same time, Pockels modulators and phase shifters will be commercialized in the SOI CMOS processes. 
LNOI will be integrated with silicon photonics processes for applications that require very high-speed modulation and low $V_\pi$, despite their longer dimensions that prohibit LSI/VLSI integration. 
\item Likewise, high-efficiency thermo-optical heaters will not be pushed out overnight. But the quest for the ideal low-power phase shifter will come up with a solution that will really enable LSI/VLSI applications. Many technologies are competing, and a clear winner is yet to emerge.  
\item Multiple layers of SiN and Si will be commonly supported in commercial foundries, and high-performance passive components (filters, delay lines) will be optimized for these SiN layers. 
\item We expect the trend for inverse design to start yielding more compact, high-performance, and robust waveguide blocks that become an integral part of PDKs. The same techniques will also fuel the performance of metamaterials and metasurfaces. 
\item The IL, bandwidth, and on-chip area for fiber-to-PIC coupling will keep improving, with the typical IL dropping to $<$ 0.5~dB.
\end{itemize}

\noindent
We also expect increased activities that open silicon photonics to a broader audience. Integrating photonic circuit design flows with (or into) electronic design automation (EDA) environments has already started, and as circuits become more complex, the codesign of photonics and electronics will become more critical.  
The scaling of photonic circuits and the convergence with electronics will also lead to greater configurability and programmability of photonic circuits, lowering the threshold for building new systems that harness the physics of light for new applications.

The space age launched the CMOS industry, the internet age launched the photonic industry, and the data age will fuel them both.

\backmatter



\bmhead{Acknowledgments}

We acknowledge Abdelrahman Afifi, Rod Augur, Jonathan Doylend, Felix Eltes, Ken Giewont, Samantha Grist, Hasitha Jayatilleka, Gordon Keeler, Matthew Mitchell, Volker Sorger, Iman Taghavi, Ming Wu, Mark Webster, and Gunay Yurtsever for technical discussions.
S.S is supported by Schmidt Science Polymath Award. S.S, L.C, B.J.S acknowledge support from the Natural Sciences and Engineering Research Council of Canada (NSERC).  J.E.B. is supported by DARPA MTO.

\bmhead{Data availability}
Data is available on reasonable request.

\bmhead{Author contributions}
S.S. led the manuscript writing and figure creation; S.S., W.B., L.C., J.E.B., M.H., R.S., and B.J.S. critically discussed the content and contributed to editing and revising the manuscript.

\bmhead{Competing interests}
S.S and L.C cofounded Dream Photonics. J.E.B cofounded Nexus Photonics and Quintessent. B.J.S cofounded Milkshake Technology. The remaining authors declare no competing interests.







\end{document}